\documentstyle[prl,aps,epsfig,twocolumn]{revtex}  

\begin{document}

\draft
\title{Quantum Information Processing with Microfabricated Optical Elements}

\author{F.B.J. Buchkremer, R. Dumke, M. Volk, T. M\"uther, G. Birkl, and W. Ertmer}

\address{Institut f\"ur Quantenoptik, Universit\"at Hannover,
Welfengarten 1, 30167 Hannover}

\date{\today}
\maketitle

\begin{abstract} 

We discuss a new direction in the field of quantum
information processing with neutral atoms. It is based on
the use of microfabricated optical elements.
With these elements versatile 
and integrated atom optical devices can be created in a compact fashion.
This approach opens the possibility to scale, 
parallelize, and miniaturize
atom optics for new investigations in fundamental research and applications 
towards quantum computing with neutral atoms. 
The exploitation of the unique features of the quantum mechanical behavior of matter waves 
and the capabilities of powerful state-of-the-art micro- and nanofabrication techniques
lend this approach a special attraction.

\end{abstract}


\narrowtext


\section {Introduction}

Following the spectacular theoretical results in the field of quantum information processing 
of recent years \cite{Bouwmeester}, 
there is now also a growing number of experimental groups working in this area. Among the 
many currently investigated approaches, which range from schemes in quantum 
optics to superconducting electronics \cite{Fortschritte}, the field of atomic physics seems to be particularly 
promising due to the experimentally achieved remarkable control of single qubit systems and the 
understanding of the relevant coherent and incoherent processes. 
While there have been successful implementions of quantum logic 
with charged atomic particles in ion traps \cite{Monroe}, 
quantum 
information schemes based on neutral atoms \cite{Hagley,Jaksch99,Brennen,Jaksch00} are an attractive alternative due to the weak 
coupling of neutral atoms to their environment. 
A further attraction of neutral atoms lies in the fact that many of the requirements for the 
implementation of quantum computation \cite{DiVincenzo} are potentially met by the newly 
emerging  
miniaturized and integrated atom optical setups. 

These miniaturized setups can be obtained by using different types of
microfabricated structures:
The trapping and guiding of neutral atoms
in microfabricated {\bf charged and current carrying} structures has been pursued
by a number of groups in recent years 
\cite{Weinstein,Hindsreview,Schmiedmayer,Haensch,Cornell,Prentiss,Engels,Hinds}. 
A new approach to generate
miniaturized and integrated atom optical systems has been recently introduced by our group 
\cite{Birkl}:
We proposed the application of
microfabricated {\bf optical} elements
(microoptical elements)
for the manipulation of atoms and atomic matter waves
with laser light.
This enables one to exploit the vast industrial and
research interest in the field of applied optics directed towards the 
development of micro-optical elements, which has already lead to a wide range
of state-of-the-art optical system applications \cite{Herzig,Sinzinger}
in this field. Applying these elements to the field of
quantum information processing, however, constitutes a novel approach. 
Together with systems based on miniaturized and microfabricated 
mechanical as well as electrostatic and magnetic
devices, the application of microoptical systems 
will launch a new field in atom optics which 
we call {\bf ATOMICS} for {\bf AT}om {\bf O}ptics with {\bf MIC}ro-{\bf S}tructures.
This field will combine the unique features of devices based on the  
quantum mechanical behavior of atomic matter waves with the 
tremendous potential of micro- and nanofabrication technology and will lead to setups that 
are also very attractive for quantum 
information processing.
 
\section {Microoptical Elements for Quantum Information Processing}

A special attraction of using microoptical elements 
lies in the fact, that 
most of the currently used techniques in atom manipulation are based on the optical
interaction with the atoms.
The use of microfabricated optical elements is therefore in many ways the canonical extension 
of the conventional optical methods into the micro-regime, so that much of the knowledge and 
experience that has been acquired in atom optics can be applied to this new regime in a 
very straightforward way. 
There are however, as we will show in the 
following, a number of additional inherent advantages in using microoptics which 
significantly enhance the applicability of atom optics and will lead to a range of new 
developments that were not achievable until now:
The use of state-of-the-art lithographic manufacturing techniques adapted from semiconductor 
processing enables the optical engineer to fabricate structures with dimensions in the 
micrometer range and submicrometer features 
with a large amount of flexibility and in a large variety of materials (glass, quartz, 
semiconductor materials, 
plastics, etc.). The flexibility of the manufacturing process allows the realization of complex 
optical elements which create light fields not achievable with standard optical components. 
Another advantage lies in the fact, that microoptics is
often produced with many identical elements fabricated in parallel on the same 
substrate, so that multiple realizations of a single conventional setup can be created
in a straightforward way. 
A further attraction of
the flexibility in 
the design and manufacturing process of microoptical components 
results from the huge potential for
integration of different elements on 
a single substrate, or, by using bonding techniques, for the integration of differently manufactured
parts into one system.
No additional restrictions arise from the small size of microoptical components 
since for most applications in atom optics, the defining parameter of an optical 
system is its numerical aperture, which for microoptical components can 
easily be as high as NA=0.5, 
due to the small focal lengths achievable.  

Among the plethora of microoptical elements that can be used for 
quantum information processing applications 
are refractive or diffractive microoptics, computer 
generated holograms,
microprisms and micromirrors, integrated waveguide optics,
near-field optics, and integrated 
techniques such as planar optics or micro-opto-electro-mechanical systems (MOEMS). 
Excellent overviews of microoptics can be found in  \cite{Herzig,Sinzinger}.
To our knowledge, of all these elements only computer generated 
holograms and phase gratings have been used in atom optics so far for guiding 
\cite{Schiffer} and trapping 
\cite{Michaud,Fournier,Ozeri} of 
atoms, while a new type of atom trap based on the near field of laser radiation has been 
proposed in \cite{Klimov}. 

In this paper we give an overview of the novel possibilities that arise for quantum information 
processing with neutral atoms if one 
employs microfabricated optical elements. 
We show how crucial components for miniaturized systems for quantum information processing 
with neutral atoms can be 
realized with microoptical elements.

\section {Experimental Setup}

\begin{figure}
   \begin{center}
   \parbox{7.5cm}{
   \epsfxsize 7.5cm
   \epsfbox{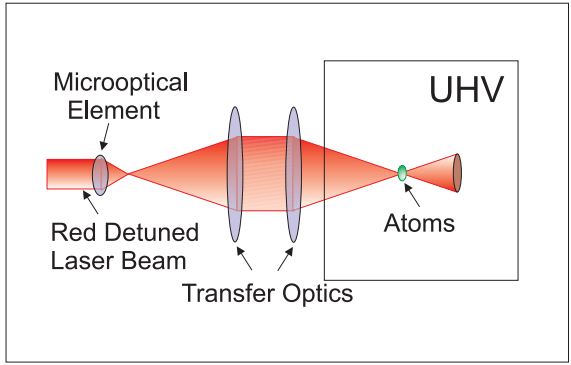}
}
   \end{center}
   \caption{Experimental setup. The light of an incoming red-detuned laser beam is 
   shaped by a microoptical component and the focal plane is imaged into the vacuum chamber. 
   The atoms are first collected in a MOT and then transfered into the 
   optical microstructure.}        
   \label{setup}
\end{figure}

The key elements in quantum information processing with neutral atoms are atom traps which 
act as storage devices for the logical qubits inscribed in the internal states of the atoms.
The experimental setup that we employ for our studies of the applicability of microoptical 
elements to generate novel trapping structures is depicted in Fig. \ref{setup}.
The focal plane of a microoptical component illuminated by a red detuned laser beam is 
transferred into the vacuum chamber with the 
help of two achromats (f = 300 mm, D = 50 mm) so that we can load the atoms into the trapping 
structures by overlapping them with a magnetooptical trap (MOT). This setup has the advantage, that 
the performance of the MOT is not disturbed by the microoptical component and that we can 
experiment with a variety of microoptical elements without the need to open the vacuum 
chamber.

\begin{figure}
   \begin{center}
   \parbox{7.5cm}{
   \epsfxsize 7.5cm
   \epsfbox{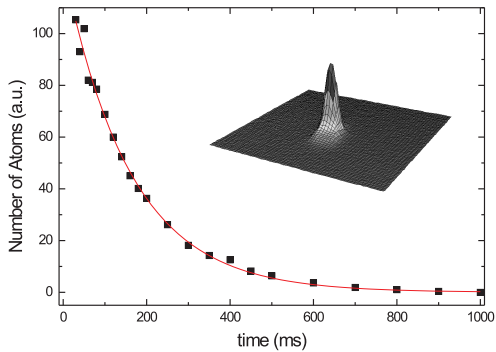}
}
   \end{center}
   \caption{Inset: Density distribution of atoms in single dipole trap obtained by using a 
   single microlens in the setup depicted in Fig. \ref{setup}. Main picture: Number of atoms 
   remaining in dipole trap as a function of time. An exponential fit to the data yields a 
   storage time of 166 ms.}        
   \label{Single-trap}
\end{figure}

A particularly simple atom trap is based on the dipole potential of 
a single focused red detuned laser beam. This trap has been first realized by Chu {\it et al.}
\cite{Chu_dipole} 
and has remained an 
important element 
ever since \cite{Grimm}. We have achieved this trapping structure by using a single lens as 
the microoptical component in Fig. \ref{setup} and have managed to trap approximately $10^5$ 
$^{85}Rb$ atoms with a lifetime of 166 ms (Fig. \ref{Single-trap}) in a dipole trap of 
potential depth 
$U_0/k_B = -1.9 mK$ (laser power P = 50 mW, detunig of 2 nm, Gaussian waist of focus $w_0$ = 
15 $\mu$m).

\section {Multiple Atom Traps}

\begin{figure}
   \begin{center}
   \parbox{7.5cm}{
   \epsfxsize 7.5cm
   \epsfbox{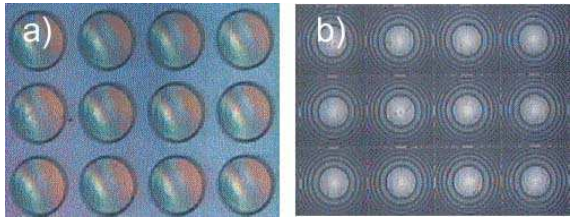}
}
   \end{center}
   \caption{Refractive (a) and diffractive (b) array of spherical microlenses.}        
   \label{Microlenses}
\end{figure}

A new approach arises from the application of one- or two-dimensional arrays of spherical 
microlenses for atom trapping (Fig. \ref{Microlenses}).    
Microlenses have
typical diameters of ten to several hundreds
of $\mu$m. Due to their short focal lengths  of typically 100$\mu$m to 1mm,
their numerical aperture can be easily as high as 0.5, resulting in foci 
whose focal size q (defined as the radius of the first minimum of the Bessel function
which results from the illumination of an individual microlens with a plane wave) can be as 
low as q=1$\mu$m for visible laser light.

\begin{figure}
   \begin{center}
   \parbox{7.5cm}{
   \epsfxsize 7.5cm
   \epsfbox{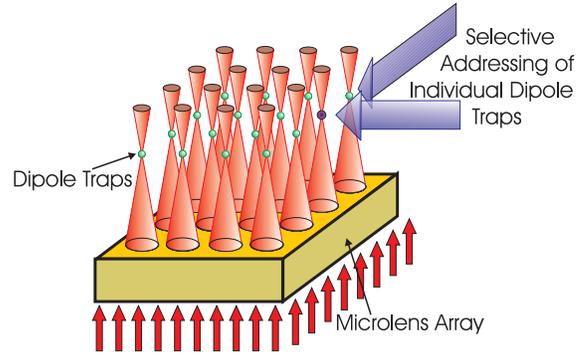}
}
   \end{center}
   
   \caption{Two-dimensional array of dipole traps created by focusing a red-detuned laser 
   beam with an array of microlenses. Due to their large separation (typically 100$\mu$m)
   individual traps can be addressed selectively, e.g. by two-photon Raman-excitation, 
   as depicted.}        
   \label{Klassiker}
\end{figure}

By focusing a single red-detuned 
laser beam extending over multiple microlenses with a spherical microlens array, we obtain 
one- or two-dimensional 
arrays of                                        
a large number of dipole traps (Fig. \ref{Klassiker}), in which we store multiple-atom 
samples \cite{Dumke}.

For frequently used atomic species and
commonly used laser sources 
one can easily obtain a
large number of atom traps
of considerable depth with rather moderate laser power. 
For typical laser parameters and 100 atom traps, the trap depth is significantly larger than the 
kinetic energy of the atoms achievable with 
Doppler cooling (0.141 mK $\times k_B$ for rubidium) \cite{Birkl}. 
The low rates 
of spontaneous scattering that are achievable with sufficiently far-detuned trapping light 
ensure long storage and coherence times as required for successful quantum information
processing, while a 
strong 
localization of the atoms
strongly suppresses heating of the atoms and makes it possible to cool the atoms to 
the ground state of the dipole potential via 
sideband cooling in all dimensions. 
For the parameters given in \cite{Birkl} the size of the atomic wavefunction reaches 
values that are significantly smaller
than 100 nm, even approaching 10 nm in many cases,
thus making microlens
arrays well suited for the generation of strongly confined 
and well localized atom samples. 

The lateral distances between the individual traps (typically 100 $\mu$m) 
make it easy to selectively detect and address the atom samples in each dipole trap.
While the natural way of addressing an individual trap
consists in sending the addressing laser beams through the corresponding microlens, 
there are also more sophisticated methods possible, e.g. with a two-photon Raman-excitation
technique as depicted in Fig. 
\ref{Klassiker}. This technique has been applied frequently to create superposition 
states in alkali atoms \cite{Kasevich_Raman} and can be used to implement single qubit 
rotations for quantum information processing. It relies on the simultaneous interaction of the 
atoms with two mutually coherent laser fields.
For a sufficiently large detuning from the single photon resonance, only the atoms in the trap 
that is addressed by both laser beams are affected by them. 

These factors open the possibility to prepare and modify quantum states 
in a controlled way ("quantum engineering") in each trap, which is a necessary ingredient for 
quantum computing. As a first and easily achievable implementation, 
single qubits associated with long-lived internal states 
can be prepared and rotated in each 
individual trap of a two-dimensional dipole trap array (Fig. \ref{Microlenses}), 
stored and later read out again. Thus, this device can serve as a quantum state register.

The manipulation of atoms with microlens arrays is extremely flexible:
It is easily possible to temporarily modify the distances between individual 
traps
if smaller or adjustable distances between traps are required. This can 
be accomplished either by using two independent microlens arrays which are laterally shifted 
with 
respect to each other or by   
illuminating a microlens array with two beams (possibly of different wavelength) under  
slightly different angles, thereby
generating two distinct sets of
dipole trap arrays. Their mutual distance can be controlled by changing 
the angle 
between the two beams. With a fast beam deflector, this
can be done in real-time during the experiment.

Due to this flexibility, setups based on microlens arrays are also well suited for the 
implementation of two-qubit gates. Considering, for example, quantum phase gates based on 
dipole-dipole interactions between atoms 
\cite{Brennen} all requirements are fulfilled in the configuration
depicted in Fig. \ref{Klassiker}. Atoms localized in neighboring traps can be first 
initialized and then be
brought close to each other with a definable separation in the single-micron range 
and for a predefined duration, in order 
to inscribe the required phase shift. Especially well suited is this configuration also for 
quantum gates based on the 
dipole-dipole interaction of low-lying Rydberg states in constant electric fields, as proposed 
in \cite{Jaksch00}.

\section{Integration}

\begin{figure}
   \begin{center}
   \parbox{9cm}{                                                                 
   \epsfxsize 9cm
   \epsfbox{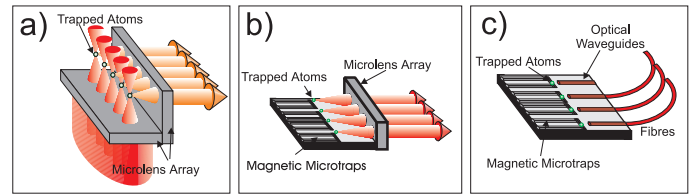}
}
   \end{center}
   \caption{Spatially resolved readout of the internal and external states
   of atoms (e.g. the state of a qubit) using microlens arrays: (a) Integration
   of two spherical microlens arrays creates a combined system of dipole
   traps and efficient detection optics.
   (b) Integration of a microlens array (for readout) with microfabricated 
   magnetic or electrostatic trapping
   structures. (c) Optical waveguides and fibres can also be integrated on the substrate.}        
   \label{readout}
\end{figure}

The huge potential for integration of microoptical components 
can be used for a large variety of further atom optical purposes. 
Due to their large numerical aperture, microoptical components can also be used
for efficient spatially resolved read-out of quantum information 
(Fig. \ref{readout}). 
In most cases the state of a qubit is recorded by exciting the atom state-selectively with 
resonant light and collecting the fluorescence light. Microoptical components can be used for 
the collection optics.
Furthermore, the optical detection of quantum states with microoptical components is
not restricted to optical trapping structures (Fig. \ref{readout} (a)):
Since the same techniques are 
applied for the fabrication of microoptical components and microstructured wires on surfaces,
microoptical components can be easily combined with the magnetic and electric structures 
of \cite{Weinstein,Hindsreview,Schmiedmayer,Haensch,Cornell,Prentiss,Engels,Hinds} 
(Fig. \ref{readout} (b)).
In addition, microfabricated atom-optical components can be integrated with optical 
fibres and waveguides
so that quantum information after being read-out 
by detection of the scattered light 
can be further processed by optical means (Fig. \ref{readout} (c)).

\begin{figure}
   \begin{center}
   \parbox{8cm}{                                                                 
   \epsfxsize 8cm
   \epsfbox{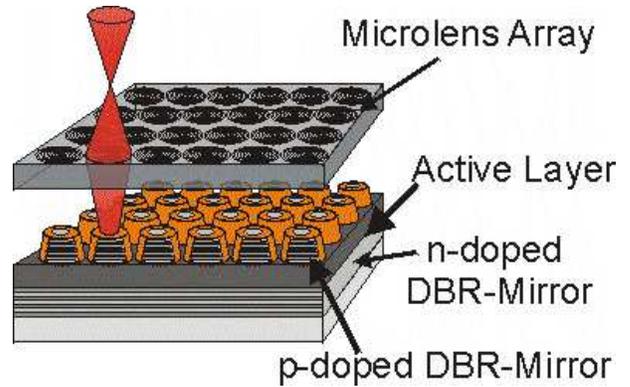}
}
   \end{center}
   \caption{Integration of microoptical components and laser sources. An array of 
   vertical cavity surface emitting lasers (VCSELs) 
   illuminates a microlens array with matched lens separation. Each trap of 
   the resulting two-dimensional 
   array of dipole traps can be individually 
   switched on and off because the individual VCSELs are selectively addressable.}        
   \label{VCSEL}
\end{figure}

Another canonical extension is given by the integration of microoptical components with 
optoelectronic devices such as semiconductor laser sources and photodiode detectors. 
In this case, the 
communication with the outside world can take place fully electronically,
with the required laser light created in situ and the optical signals converted back to 
electrical signals on the same integrated structure.
Fig. \ref{VCSEL} illustrates a simple but powerful configuration based on this approach. 
The depicted setup utilizes vertical cavity 
surface emitting lasers (VCSELs) \cite{Sinzinger,Iga,Jewell} which are directly mounted onto
a two-dimensional microlens array. 
VCSELs emit circular symmetrical non-astigmatic beams while their structure is optimized for the 
lithographic fabrication of densely packed 2D arrays. Since VCSELs generally emit 
a single longitudinal mode at a wavelength which can be tuned by changing
the current, the light coming from a 
two-dimensional array of VCSELs can be directly focused by the mirolens array
to create a two-dimensional array of stable dipole traps (Fig. \ref{VCSEL}).
With VCSELs
it becomes now possible to selectively switch on and off individual traps or to selectively 
change their potential depth resulting in further flexibility in the atom manipulation.

\section {Conclusion}

In this paper we have discussed the new research direction of using microfabricated optical 
elements for quantum information processing with neutral atoms.
This application 
hugely benefits from the many inherent advantages of microoptical 
components.
Specifically, an approach based on microoptical systems 
addresses two of the most important requirements for the
technological implementation of quantum information processing: parallelization and scalability.
In addition, the possibility to selectively address individual qubits 
is essential for most schemes proposed for quantum 
computing with 
neutral atoms. 
 
Thus, all steps required for quantum information processing with neutral atoms -
i.e. the preparation, manipulation and storage of qubits, entanglement and gate operations as 
well as the efficient 
read-out of quantum information - 
can be performed using microfabricated optical elements.

\section {Acknowledgements}

This work is supported by the program ACQUIRE (IST-1999-11055) of the European Commission 
as well as the 
SFB 407 and the {\it Schwerpunktprogramm Quanten-Informationsverarbeitung} of the {\it Deutsche 
Forschungsgemeinschaft}.


 \newpage
%
%

%
%

%
%

%
%
%

%
%

%
%

%
%

%


\end{document}